# Iron Partitioning between Ferropericlase and Bridgmanite in the Earth's Lower Mantle

Shenzhen Xu[1], Jung-Fu Lin[2,3], and Dane Morgan[1,4]

[1]Materials Science Program, University of Wisconsin- Madison, Madison, WI, USA

[2]Department of Geological Sciences, Jackson School of Geoscience
The University of Texas at Austin, Austin, TX, USA

[3]Center for High Pressure Science and Technology Advanced Research (HPSTAR)
Shanghai, China

[4]Department of Materials Science and Engineering
University of Wisconsin – Madison, Madison, WI, USA

Corresponding author: Dane Morgan (ddmorgan@wisc.edu)

**Key Points:**

1. We build an integrated thermodynamic model of Fe and Al chemistry for lower mantle conditions.
2. We find a new stratified picture of the Fe speciation profile with respect to pressures.
3. The Fe partitioning between ferropericlase and bridgmanite is dominated by Fp $Fe^{2+}$ spin transition.

**Abstract:**

Earth's lower mantle is generally believed to be seismically and chemically homogeneous because most of the key seismic parameters can be explained using a simplified mineralogical model at expected pressure-temperature conditions. However, recent high-resolution tomographic images have revealed seismic and chemical stratification in the middle-to-lower parts of the lower mantle. Thus far, the mechanism for the compositional stratification and seismic inhomogeneity, especially their relationship with the speciation of iron in the lower mantle, remains poorly understood. We have built a complete and integrated thermodynamic model of iron and aluminum chemistry for lower mantle conditions, and from this model has emerged a stratified picture of the valence, spin, and composition profile in the lower mantle. Within this picture the lower mantle has an upper region with $Fe^{3+}$ enriched bridgmanite with high-spin ferropericlase and metallic Fe, and a lower region with low-spin, iron-enriched ferropericlase coexisting with iron-depleted bridgmanite and almost no metallic Fe. The transition between the regions occurs at a depth of around 1600km and is driven by the spin transition in ferropericlase, which significantly changes the iron partitioning and speciation to one that favors $Fe^{2+}$ in ferropericlase and suppresses $Fe^{3+}$ and metallic iron formation. These changes lead to lowered bulk sound velocity by 3-4% around the mid-lower mantle and enhanced density by ~1% toward the lowermost mantle. The predicted chemically and seismically stratified lower mantle differs dramatically from the traditional homogeneous model.

**1. Introduction:**

Recent mineral physics studies have revealed the complex nature of the spin and valence states of iron in ferropericlase (Fp) and bridgmanite (Pv) at high pressure and temperature (*P-T*). The spin crossover of $Fe^{2+}$ in Fp occurs over a wide *P-T* range corresponding to approximately 1100km to 1900km in depth [*Badro et al.*, 2003; *Goncharov et al.*, 2006; *J. F. Lin et al.*, 2005; *Persson et al.*, 2006; *Sturhahn et al.*, 2005; *Tsuchiya et al.*, 2006; *Vilella et al.*, 2015], whereas the $Fe^{3+}$ in the octahedral site (B site) of Pv undergoes the transition from about 1000km to 1500km in depth [*Krystle Catalli et al.*, 2011a; *Hsu et al.*, 2011; *Hsu et al.*, 2012], both depending somewhat on temperature. Of particular interest to our understanding of deep-mantle seismology are the effects of the spin transition on the density and sound velocity profiles in lower mantle Fp [*Cammarano et al.*, 2010; *Crowhurst et al.*, 2008; *Marquardt et al.*, 2009; *Wu and Wentzcovitch*, 2014], where abnormal velocity softening within the transition has been reported and the full low-spin state has been found to exhibit distinct elastic behaviors from its full high-spin counterpart. Also of interest is the variation of Fe valence states ($Fe^{2+}$, $Fe^{3+}$, $Fe^{0}$) with respect to depth because this variation determines both the amount of Fe that undergoes each type of spin transition and the amount of metallic Fe ($Fe^{0}$), which has a significant influence on our understanding of the partial melting in the mantle [*Fukai and Suzuki*, 1986] and the Earth's core formation [*Halliday and Lee*, 1999]. In addition, the Fe partitioning between Pv and Fp ($K_D^{Pv-Fp}$=(Fe/Mg)$_{Pv}$/(Fe/Mg)$_{Fp}$) is also critical to the interpretation of the deep-Earth geochemistry, because changes in partitioning can lead to a chemically stratified lower mantle that may behave distinctly from a traditional homogenous model. While all these separate aspects of spin, valence, and partitioning are important, their impact on the lower mantle can only be properly determined when their couplings are taken into account through an integrated picture of Fe chemistry under lower mantle conditions. The Fe chemistry is expected to couple to multiple interacting aspects of the lower mantle environment, including *P-T*, spin transitions, oxygen fugacity, and aluminum substitution. Understanding and quantifying such couplings requires an integrated thermodynamic model of Fe chemistry that incorporates all of these environmental factors.

Using a pyrolitic composition, iron has been shown experimentally to partition into Pv and Fp with a $K_D^{Pv-Fp} \approx 0.5$ ($K_D^{Pv-Fp}$ is defined as $[Fe^{2+}+Fe^{3+}]_{Pv}/[Mg^{2+}]_{Pv})/([Fe^{2+}]_{Fp}/[Mg^{2+}]_{Fp})$ [*Irifune et al.*, 2010; *Nakajima et al.*, 2012]. However, $K_D^{Pv-Fp}$ was observed to significantly increase to almost 0.9 at approximately 28 GPa (or ~750-km depth) [*Irifune et al.*, 2010]. The change in $K_D^{Pv-Fp}$ has been attributed to the formation of $Fe^{3+}$-rich Pv promoted by the charge-coupled substitution of $Al^{3+}$ and $Fe^{3+}$ [*Frost et al.*, 2004; *Irifune et al.*, 2010]. On the other hand, self-disproportionation of $Fe^{2+}$ into $Fe^0$ and $Fe^{3+}$ has also been used to explain the observations of the high $Fe^{3+}$ content in Pv and the co-existence of metallic iron [*Frost et al.*, 2004]. At *P-T* conditions below the mid-lower mantle, experimental studies have also shown that $K_D^{Pv-Fp}$ decreases to approximately 0.5 in the pyrolitic composition or to 0.2 in the olivine composition [*Auzende et al.*, 2008; *Irifune*, 1994; *Irifune et al.*, 2010; *Kesson et al.*, 1998; *Kobayashi et al.*, 2005; *Sakai et al.*, 2009; *Sinmyo et al.*, 2008a; *Wood*, 2000]. A recent theoretical work of J. Muir and J. Brodholt [*Muir and Brodholt*, 2016] studied the ferrous iron partitioning between Pv and Fp and predicted that $Fe^{2+}$ prefers Fp rather than Pv. Their calculations showed a convex-shape feature of the $K_D^{Pv-Fp}(P)$ profile with a peak ($K_D^{Pv-Fp}$ ~0.25) at 50GPa and gradually decreasing to 0.05 at higher pressures. However, they only calculated the ferrous iron $Fe^{2+}$ partitioning in the two phases, which is $(Mg_{1-x},Fe^{2+}_x)O + MgSiO_3 \Leftrightarrow (Mg_{1-y},Fe^{2+}_y)SiO_3 + MgO$. The important impurity species $Al^{3+}$ and $Fe^{3+}$ in the Pv phase, the possible formation of metallic Fe phase and the influence of the surrounding oxygen fugacity (*f*O$_2$) conditions are not considered in their thermodynamic model. Thus far, the relationship of Fe partitioning with Fe valence and spin states at lower mantle relevant *P-T*, composition (*X*), and *f*O$_2$ conditions remain largely unexplored. Here we have developed an integrated and validated *ab-initio* and empirical fitting based thermodynamic model to explicitly predict the partitioning behavior of Fe in lower mantle Pv and Fp over a wide range of likely *P-T-X-f*O$_2$ conditions in pyrolite and olivine compositional models. We have used density functional theory (DFT) calculations with the HSE06 hybrid exchange-correlation functional [*Heyd et al.*, 2003] for the relevant enthalpies. The equilibrium iron valence, spin states, and occupation in the sublattices of Fp and Pv are determined by minimizing the total Gibbs free energy of the system (See Methods section 2 for details). Since it is well known that the variation in the oxygen fugacity ($fO_2$) affects Fe valence states in the upper-mantle minerals it is important to consider its possible impact on the lower-mantle bridgmanite, and therefore here we have considered two different scenarios in our modeling: a closed system with no exchange of free oxygen with surrounding materials where $Fe^{3+}$ is produced through the charge disproportionation (chg. disp.) reaction ($3Fe^{2+} \rightarrow 2Fe^{3+}+Fe^0$) [*Frost et al.*, 2004; *Xu et al.*, 2015; *Zhang and Oganov*, 2006] and an open system with free oxygen buffered through oxygen exchange with surrounding materials in the system (See Methods section 2 for details). These modeling approaches allow us to investigate the iron chemistry under lower mantle conditions (closed system scenario) and to compare with experimental observations where the $fO_2$ is usually buffered by certain capsules (open system scenario), typically resulting in relatively oxidizing conditions compared to the lower mantle [*Irifune et al.*, 2010]. The effects of $Al^{3+}$ incorporation are investigated via the integration of the $Al^{3+}$ substitution energetics in the pyrolite system, and comparison with the olivine system yields understanding of the discrepancy between Al-bearing and Al-free systems of the reported $K_D$ values in previous experimental reports [*Auzende et al.*, 2008; *Irifune*, 1994; *Irifune et al.*, 2010; *Kesson et al.*, 1998; *Kobayashi et al.*, 2005; *Sakai et al.*, 2009; *Sinmyo et al.*, 2008a; *Wood*, 2000].

## 2. Results and Discussion:

Our *ab-initio* and empirical fitting based thermodynamic model predicts that a gradual spin transition occurs for $Fe^{2+}$ in Fp from approximately 40 GPa to 80 GPa and for B-site $Fe^{3+}$ in Pv from approximately 30 GPa to 50 GPa (Fig. 1(a)), both of which are along an expected geotherm [*Brown and Shankland*, 1981]. The predicted fractions of different Fe species in the pyrolitic compositional system show that high-spin $Fe^{2+}$ is the predominant form of Fe in Fp and Pv around the top of the lower mantle and it should be noted that $Fe^{3+}$ is produced only through the chg. disp. reaction in Pv. Our model predicts the formation of metallic Fe from the top to the mid-lower mantle. Once sufficient $Al^{3+}$ is dissolved into Pv, most B-site $Fe^{3+}$ is driven to the A-site (Fig. 1(b)) (details of the Al content in Pv are discussed in Method section 1, and the driving force for A-site occupancy is discussed in Supplemental Information (SI) section 2). The loss of B-site $Fe^{3+}$ implies that the spin transition of Pv B-site $Fe^{3+}$ will not have a significant impact on the physical and chemical properties of the lower mantle [*Hsu et al.*, 2012]. Furthermore, $Fe^{2+}$ preferentially partitions into Fp across its spin transition such that Pv is depleted of iron towards the lowermost-mantle. This spin-induced iron partitioning suppresses the aforementioned chg. disp. reaction, significantly decreasing the presence of the metallic iron in the lower parts of the lower mantle. Therefore the formation of metallic Fe is predicted to occur only from the top to the mid-lower mantle.

The partitioning coefficient ($K_D^{Pv-Fp}$) is calculated using the predicted iron contents in Fp and Pv for expected geotherm *P-T* conditions of the pyrolitic lower mantle (Fig. 2(a), and $\ln(K_D^{Pv-Fp})$ is given in SI Fig. S6). Our predicted $K_D$ curve shows a convex shape starting with an initial value of 0.2 at 22 GPa, a peak value of 0.34 at 40 GPa, and a continuous decrease between 40 GP and 120 GPa reaching a minimum value of almost zero. At the topmost lower mantle, $Fe^{2+}$ is more energetically favorable to partition into Fp. As the *P-T* increases up to 40 GPa, the incorporation of $Al^{3+}$ into Pv through the dissolution of majorite drives both high-spin $Fe^{2+}$ and $Fe^{3+}$ into the A-site of the Pv lattice. Another reason causing more iron partitioning into Pv from 20 to 40GPa is that the majority of Fp $Fe^{2+}$ is in the high-spin state, where its enthalpy becomes relatively higher as pressure increases as compared to Pv $Fe^{2+}$ (see SI section 2.3 for detailed discussion). Compared with the $K_D$ curve without the spin transition in Fp that continuously increases with increasing pressure (Fig. 2(a)), it is evident that the dramatic decrease in the $K_D$ curve starting at approximately 40 GPa is caused by the spin transition in Fp. The volume collapse and the associated reduction in the Gibbs free energy make the low-spin $Fe^{2+}$ with a smaller volume more energetically favorable in Fp, as opposed to staying in Pv. As the spin crossover continues to occur with increasing depth, the $K_D$ value below the spin transition zone reaches almost zero, and the majority of the iron partitions into Fp at the lowermost mantle. This partitioning also destabilizes metallic iron in the system, reducing it to less than 1% of the Fe towards the deeper lower mantle. The Fp $Fe^{2+}$ spin transition governing Fe partitioning profile shown in our calculation matches very well with the results in the recent theoretical work of J. Muir and J. Brodholt [*Muir and Brodholt*, 2016]. The same convex-shape feature of the Fe partitioning coefficient with respect to pressures was reported in their work (Fig. 7 in the paper [*Muir and Brodholt*, 2016]). The peak position was 50GPa in their calculation, very close to the 40GPa peak position in our results (Fig. 2(a)). For the quantitative comparison, the peak value of Fe partitioning coefficient was about 0.25 in their results, while we find a peak value about 0.35 in our calculation. This small quantitative difference may come from the fact that Muir and Brodholt did not include the $Fe^{3+}$ and $Al^{3+}$ in the Pv phase in their model. As the existence of $Fe^{3+}$ and $Al^{3+}$ increases the Fe content in the Pv phase, their inclusion in our model is expected to yield a relatively higher Fe partitioning coefficient $K_D^{Pv-Fp}$. In addition, Muir and Brodholt based their DFT calculations on the generalized gradient approximation with the Hubbard U corrections (GGA+U) while we used the HSE06 hybrid functional in our work, which is generally believed to be a more consistent and accurate method. However, our results are consistent with their findings both qualitatively and quantitatively, lending support to both models.

To understand the consequences of Fp $Fe^{2+}$ spin transition and Fe partitioning on the seismic profiles of the Earth's lower mantle, we have also modeled the mass density ($\rho$) and bulk sound velocity ($V_\Phi$) profiles of a representative pyrolitic lower mantle composition along an expected mantle geotherm [*Brown and Shankland*, 1981] (calculation details are shown in SI section 4.2 [*Tange et al.*, 2012; *Tange et al.*, 2009] *[Dubrovinsky et al., 2000; Lu et al., 2013; Murakami et al., 2012; Ricolleau et al., 2010; Shukla et al., 2015; van Westrenen et al., 2005; X Wang et al., 2015; Wu and Wentzcovitch, 2014]*) and compared them with the density and bulk velocity without the $Fe^{2+}$ spin transition in Fp (Fig. 2(b)). Here the bulk sound velocity is defined as $V_\Phi = (K_S/\rho)^{1/2}$, where $K_S$ is the adiabatic bulk modulus and values are taken along the geotherm. We only include Pv, Fp and metallic Fe in our model as they occupy more than 90% of the volume of the lower mantle (SI section 4.2). The starting zero-pressure relative volumetric ratio of Pv over Fp in our pyrolitic model is $V$((Al, Fe)-bearing Pv):$V$(Fp) = 3.5:1 [*Irifune et al.*, 2010]. Our calculated density and velocity profiles display continuous increase with increasing depth without any abrupt changes. However, compared with the calculated profiles without the spin transition, we do notice lowered bulk sound velocity around mid-lower mantle by 3~4% and enhanced density by 1% towards the bottom of the lower mantle. These changes can be explained by the spin transition of $Fe^{2+}$ in Fp. As the low-spin $Fe^{2+}$ in Fp has a relatively smaller volume than its high-spin counterpart, the volume of Fp will decrease when the spin transition occurs, increasing the density of the system near the bottom of the lower mantle. This volume collapse also causes a softened bulk modulus in the spin transition region (40~80GPa) leading to a lowered bulk sound velocity ($V_\phi = (K_S/\rho)^{1/2}$) for the system due to the spin transition. The predicted profiles also show a lowered bulk sound velocity by up to 2% with respect to the PREM model (see SI section 3.4 [*Dziewonski and Anderson*, 1981]).

Comparison of our theoretical predictions in the oxidation reaction model at experimental $P$-$T$-$fO_2$ conditions with previous experimental results [*Auzende et al.*, 2008; *Irifune*, 1994; *Irifune et al.*, 2010; *Kesson et al.*, 1998; *Kobayashi et al.*, 2005; *Sakai et al.*, 2009; *Sinmyo et al.*, 2008a; *Wood*, 2000] for the iron partitioning shows distinctions between the pyrolite composition containing $Al^{3+}$ and the olivine bulk composition without $Al^{3+}$. We use a representative $fO_2$ range of log$fO_2$[diamond-carbonate] ≤ log$fO_2$ ≤ log$fO_2$[diamond-carbonate] + 2, as discussed in SI section 3.2 [*Irifune et al.*, 2010; *Pownceby and O'Neil*, 1994; *Stagno and Frost*, 2010; *Stagno et al.*, 2015]. Here the symbol $fO_2$[X] represents the fugacity of material X under the given $P$-$T$ conditions. **For the notation of $fO_2$[diamond-carbonate], diamond-carbonate represents the equilibrium described by the reaction C (diamond) + $O^{2-}$ (mineral/melt) + $O_2$ <=> $CO_3^{2-}$ (mineral/melt) [*Stagno and Frost*, 2010]. The equilibrium of carbon (diamond) coexisting with carbonates happens in the (lower-mantle relevant) high-pressure condition. The above reaction sets the $fO_2$ of the diamond capsule. We assume that within the lower-mantle relevant $P$-$T$ of our thermodynamic model, the $fO_2$[diamond-carbonate] is always 4 orders of magnitude lower than the Re-ReO$_2$ capsule $fO_2$ [*Nakajima et al.*, 2012; *Pownceby and O'Neil*, 1994; *Stagno and Frost*, 2010; *Xu et al.*, 2015]. The Re-ReO$_2$ capsule $fO_2$ values as a function of $P$-$T$ are explicitly shown in our previous work [*Xu et al.*, 2015]** Although both of the pyrolitic composition curve and the olivine-composition curve show convex $K_D$ behavior with respect to depth, the magnitude of the $K_D$ in the olivine model is a smaller by a factor of 3~4 than that in the pyrolite model (Fig. 3). Under the $fO_2$ range studied no chg. disp. occurs and more $Fe^{3+}$ is formed and incorporated into Pv compared to under lower mantle $f$O$_2$, leading to a higher $K_D$ than the chg. disp. reaction model. **The contents of different Fe species (valence and spin states) in the pyrolitic compositional model under the $f$O$_2$[diamond-carbonate] condition are shown in SI Section 3.3 Fig. S7.** This $f$O$_2$ effect explains why the pyrolitic $K_D$ profile in Fig. 3 (oxidation reaction model) is higher than that in Fig. 2(a) (chg. disp. reaction) and demonstrates the importance of maintaining realistic lower mantle $f$O$_2$ when measuring $K_D$. The difference in the magnitude between the pyrolite and olivine compositional models can be explained by the existence of Al$^{3+}$, which enhances the Fe$^{3+}$ content in Pv in the pyrolite model due to the formation of the stable Fe$^{3+}$-Al$^{3+}$ pair. As there is no Al$^{3+}$ in the olivine model, the $K_D$ values are predicted to remain at 0.1-0.2. For the pyrolite composition, the $K_D$ profiles from experiments [*Irifune*, 1994; *Irifune et al.*, 2010; *Kesson et al.*, 1998; *Murakami et al.*, 2005; *Wood*, 2000] and our simulations both exhibit a convex shape. However, the experimental results show a sharp decrease at approximately 40GPa and a flattening $K_D$ value of around 0.5 at higher pressures, which differ from our predicted gradual decrease of $K_D$ (Figs. 2(a) and 3). The gradual decrease in our calculated $K_D$ values is due to the broad spin crossover of Fp Fe$^{2+}$ at high $P$-$T$ conditions of the lower mantle [*Crowhurst et al.*, 2008; *J. F. Lin et al.*, 2005; *Tsuchiya et al.*, 2006]. Future high $P$-$T$ experimental data with lower uncertainties are needed to explain the discrepancies (See SI section 3.3 for further discussion of the discrepancies [*Ammann et al.*, 2010; *Irifune*, 1994; *Irifune et al.*, 2010; *Murakami et al.*, 2005; *Xu et al.*, 2015]).

Recent two experimental works about the Fe partitioning behavior in a pyrolitic compositional sample under lower mantle relevant P-T condition showed a large $K_D$ increase at about 100GPa [*Prescher et al.*, 2014; *Sinmyo and Hirose,* 2013]. The explanation proposed in the work [*Prescher et al.*, 2014] for the $K_D$ increase is that the Fe$^{2+}$ undergoes a spin transition from the intermediate spin (IS) to the low spin state in Pv, which is inconsistent with the widely held belief that Fe$^{2+}$ is always in the HS state in Pv throughout the lower mantle pressure range [*Bengtson et al.*, 2008; *Grocholski et al.*, 2009; *Hsu et al.*, 2010a; *Jackson*, 2005; *J.F. Lin et al.*, 2012; *Stackhouse et al.*, 2007]. Moreover, even if we assume this spin transition occurs, the same $K_D$ increase behavior would then be expected happen for the Al-free case as well. However, many previous experimental works failed to observe this abnormally high $K_D$ around 100GPa[*Auzende et al.*, 2008; *Nakajima et al.*, 2012; *Sakai et al.*, 2009]. The explanation for this $K_D$ increase proposed in the work [*Sinmyo* and *Hirose,* 2013] is the B-site $Fe^{3+}$ spin transition in Pv. However, we would argue that this is unlikely to be the mechanism causing the $K_D$ increase, because the population of $Fe^{3+}$ in B-site in the Al-bearing Pv is very small [*Hsu et al.*, 2012; *Zhang* and *Oganov*, 2006] (also discussed in SI Section 2.1 in our work) and this spin transition region reported in many previous works is from the upper to the mid lower mantle [*K. Catalli et al.*, 2010; *Hsu et al.*, 2011; *J.F. Lin et al.*, 2012; *Xu et al.*, 2015]. Specifically, the experimental work of [*J.F. Lin et al.,* 2012] reported that the B-site $Fe^{3+}$ spin transition is at about 25GPa based on their quadrupole splitting (QS) measurement. The simulation work of [*Hsu et al.*, 2011] predicted that the gradual B-site $Fe^{3+}$ spin transition is from 40GPa to 70GPa at *T*=2000K. So it's less likely for this B-site $Fe^{3+}$ spin transition to have such a significant impact on Fe partitioning above 100GPa. Moreover, some previous experimental works also showed that there is no $K_D$ increase around or above 100GPa [*Irifune et al.,* 2010; *Kesson et al.*, 1998]. Even in the paper [*Prescher et al.,* 2014], they showed that $K_D$ drops back to 0.5 at about 130GPa. If this abnormal $K_D$ increase is indeed caused by the B-site $Fe^{3+}$ spin transition, $K_D$ would stay at the high value down to the bottom of the lower mantle, which is inconsistent with the observation of [*Prescher et al.*, 2014], We also note that the explanations for the abnormal $K_D$ increase in the experimental works [*Prescher et al.,* 2014]and [*Sinmyo* and *Hirose*, 2013] are inconsistent with each other. Overall, the collected experimental results and efforts to explain them show that there is still significant uncertainty for the experimental $K_D$ at high pressure (>100GPa). More work is needed to robustly determine $K_D$ under these extreme conditions.

## 3. Conclusions

Our calculations have demonstrated that the Earth's lower mantle can be separated into distinct layers that are controlled by the spin crossover in Fp and the $Al^{3+}$ substitution in Pv (summarized in Fig. 4). The top layer is characterized by $Fe^{3+}$-enriched Pv coexisting with high-spin Fp and metallic iron, while the bottom layer at depths approximately below about 2000 km exhibits Fe predominantly in Fp and almost no metallic Fe. The transition between the two layers occurs through the Fp $Fe^{2+}$ spin transition zone (Fig. 4). The volume collapse of Fp caused by the spin crossover also leads to a lowered bulk sound velocity around the mid-lower mantle (Fig. 2(b)). These changes suggest that there are significant iron-chemistry-induced stratification of the lower mantle, in contrast to the traditional view of chemical and seismic homogeneity in the lower mantle.

## 4. Methods
### 4.1. Composition models of our thermodynamic system
#### 4.1.1. Pyrolitic lower-mantle composition

Earth's lower mantle is proposed to consist of pyrolite which contains approximately one-third basalt and two-third peridotite [*Ringwood*, 1966]. In the pyrolitic compositional model, the mineralogy of the lower mantle is mainly made of approximately 70% bridgmanite ($(Mg,Fe)(Si,Al)O_3$; Pv), 20% ferropericlase ($(Mg,Fe)O$; Fp), and 10% calcium silicate perovskite ($CaSiO_3$; Ca-Pv) [*Irifune et al.*, 2010], where all percentages are by volume. Since Ca-Pv likely does not incorporate significant Fe in its lattice, we have only considered Pv and Fp phases in the lower-mantle system with a molar ratio $[(Mg,Fe)(Si,Al)O_3] : [(Mg,Fe)O] = 1 : 0.65$ corresponding to the volumetric ratio $V[(Mg,Fe)(Si,Al)O_3] : V[(Mg,Fe)O] = 3.5 : 1$. The $Al^{3+}$ cation is considered to be mainly incorporated into Pv in the lower mantle. Previous studies have shown that the $Al^{3+}$ content in Pv increases with increasing depth at the topmost lower mantle as a result of majorite dissolution into Pv [*Irifune*, 1994; *Irifune et al.*, 2010]. The depth-dependent $Al^{3+}$ content in Pv has been considered in our calculations in order to understand its incorporation mechanism and effects on the iron partitioning between Pv and Fp phases. As the starting composition of the pyrolite system in our calculations, $Al^{3+}$ content in Pv increases from 0.06 to 0.09 per 3 oxygen atoms from 20GPa to 30GPa, and stays at 0.09 per 3 oxygen atoms from 30GPa to 120GPa, and Fe content is 0.1 per 3 oxygen atoms in Pv and 0.12 per oxygen atom in Fp from 20GPa to 120 GPa. The values are based on experimental measurements of Al and Fe content from the previous reference [*Irifune et al.*, 2010].

**4.1.2. Olivine compositional model**

In the olivine compositional model we treat the composition as equal to that of olivine. olivine, the most abundant mineral in the upper mantle, transforms to Pv and Fp in the lower-mantle P-T conditions. The Fe partitioning coefficient ($K_D$) between Pv and Fp phases with a bulk olivine composition in the lower mantle is also considered in our calculations. Since this system contains a negligible amount of $Al^{3+}$, it is also used to understand how the substitution of $Al^{3+}$ in Pv affects the iron partitioning compared to pyrolitic composition case. The typical Fe content in olivine is about 0.1 Fe per Mg-site [*Auzende et al.*, 2008; *Sinmyo et al.*, 2008b]. For the starting composition of the olivine system in our calculation, we have used San Carlos olivine with a representative chemical formula $(Mg_{0.9}Fe_{0.1})_2SiO_4$.

**4.2. Thermodynamic modeling of the Pv+Fp system**

To understand the Fe partitioning the lower mantle under relevant experimental conditions, we have modeled a bulk composition of both pyrolite and olivine, as described in Method section 1.1 and 1.2. In the model, the total amount of iron is fixed throughout all lower mantle conditions. Although the thermodynamic equilibrium state depends on overall stoichiometry and chemical potentials, the equilibrium state is independent of how we approach it. However, for clarity we consider the system to be equilibrating from an initial state with specific spin and site occupancies consistent with the overall stoichiometric constraints. We take our initial state as high-spin (HS) $Fe^{2+}$ in both Pv and Fp, in which the HS $Fe^{2+}$ occupies the A site in Pv and substitutes for the Mg ions in Fp at the relevant *P-T* of the lower mantle [*Bengtson et al.*, 2008; *Grocholski et al.*, 2009; *Zhang and Oganov*, 2006]. We also assume that $Al^{3+}$ enters into the Pv lattice through the charge-coupled substitution in which $Al^{3+}$-$Al^{3+}$ occupy the A site and B site jointly [*Brodholt*, 2000]. These initial states are allowed to vary in our calculations in order to reach thermodynamic equilibrium, in which variations are considered for Fe HS and LS states, Fe occupancy of A and B sites in Pv, $Fe^{2+}$ and $Fe^{3+}$ valence states, and partitioning of Fe between Fp and Pv.

In order to find the final equilibrium state of the system at a given *P-T* condition, various potential valence and spin states and site occupancies of iron in Pv and Fp phases have been considered in our calculations, as well as two methods of treating the oxygen availability (oxygen fugacity). Specifically, we have considered an oxidation reaction model (Eq. M1) and a chg. disp. reaction model (Eq. M2) in our thermodynamic modeling. Variations in the oxygen fugacity ($fO_2$) are also considered in the oxidation reaction model. Consideration of these different reactions helps us to understand both laboratory experimental results and behavior in the deep mantle, which can have significantly different $fO_2$ values.

The oxidation reaction model is appropriate for modeling laboratory experiments. In the laboratory, the data were typically derived from chemical analyses of quenched samples originally equilibrated at high temperatures in systems with capsules that effectively buffer the $fO_2$. Depending on the type of the capsule used (e.g., metal, MgO, diamond) and potentially the kinetics in the experiment, the $fO_2$ of the system can vary significantly, making the interpretation of experimental results more difficult [*Campbell et al.*, 2009; *Frost et al.*, 2004; *Irifune et al.*, 2010; *Lauterbach et al.*, 2000; *Nakajima et al.*, 2012; *Xu et al.*, 2015]. Under equilibrium with the $fO_2$ set by capsules used in most experimental conditions, metallic Fe cannot form, as it will be oxidized, and there is excess oxygen available to oxidize $Fe^{2+}$ to $Fe^{3+}$ [*Xu et al.*, 2015]. This explains why we don't have metallic Fe ($Fe^0$) in the oxidation model as shown in the equation M1. In contrast, it is believed that there is no excess oxygen available in the lower mantle, and any $Fe^{3+}$ that forms is created by the chg. disp. reaction ($3Fe^{2+} \rightarrow 2Fe^{3+}+Fe^0$) [*Frost et al.*, 2004; *McCammon*, 1997; *Xu et al.*, 2015].

The chg. disp. reaction model is appropriate for modeling lower mantle relevant conditions. In this model, we have assumed that the lower mantle is not chemically reacting with the surrounding layers of the upper mantle and the outer core such that there is neither external free oxygen gas nor any significant amount of other oxidizing agents to oxidize $Fe^{2+}$ available. In such a scenario, the only possible mechanism to produce $Fe^{3+}$ is via the chg. disp. reaction.

In the oxidation reaction modeling we consider the following equilibration from our initial state.

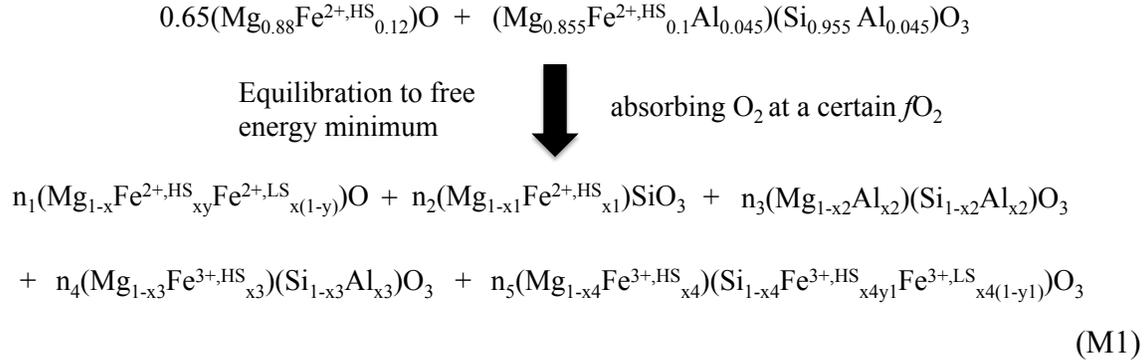

$$0.65(Mg_{0.88}Fe^{2+,HS}_{0.12})O + (Mg_{0.855}Fe^{2+,HS}_{0.1}Al_{0.045})(Si_{0.955}Al_{0.045})O_3$$

Equilibration to free energy minimum ↓ absorbing $O_2$ at a certain $fO_2$

$$n_1(Mg_{1-x}Fe^{2+,HS}_{xy}Fe^{2+,LS}_{x(1-y)})O + n_2(Mg_{1-x1}Fe^{2+,HS}_{x1})SiO_3 + n_3(Mg_{1-x2}Al_{x2})(Si_{1-x2}Al_{x2})O_3$$

$$+ n_4(Mg_{1-x3}Fe^{3+,HS}_{x3})(Si_{1-x3}Al_{x3})O_3 + n_5(Mg_{1-x4}Fe^{3+,HS}_{x4})(Si_{1-x4}Fe^{3+,HS}_{x4y1}Fe^{3+,LS}_{x4(1-y1)})O_3$$

(M1)

In the chg. disp. reaction modeling we consider the following equilibration from our initial state.

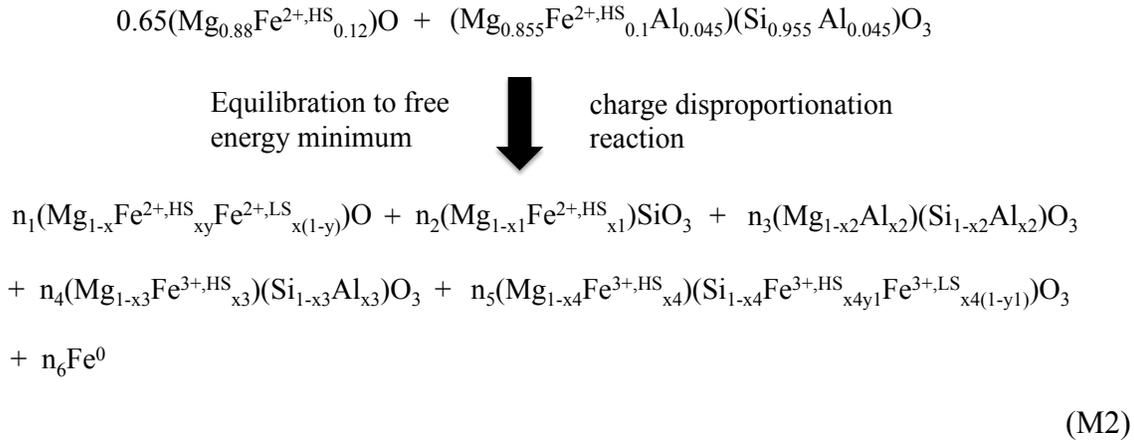

$$0.65(Mg_{0.88}Fe^{2+,HS}_{0.12})O + (Mg_{0.855}Fe^{2+,HS}_{0.1}Al_{0.045})(Si_{0.955}Al_{0.045})O_3$$

Equilibration to free energy minimum ↓ charge disproportionation reaction

$$n_1(Mg_{1-x}Fe^{2+,HS}_{xy}Fe^{2+,LS}_{x(1-y)})O + n_2(Mg_{1-x1}Fe^{2+,HS}_{x1})SiO_3 + n_3(Mg_{1-x2}Al_{x2})(Si_{1-x2}Al_{x2})O_3$$

$$+ n_4(Mg_{1-x3}Fe^{3+,HS}_{x3})(Si_{1-x3}Al_{x3})O_3 + n_5(Mg_{1-x4}Fe^{3+,HS}_{x4})(Si_{1-x4}Fe^{3+,HS}_{x4y1}Fe^{3+,LS}_{x4(1-y1)})O_3$$

$$+ n_6Fe^0$$

(M2)

For the oxidation reaction model, the grand potential of the system is minimized as a function of $P$, $T$, and the chemical potential of oxygen $\mu(O_2)$ such that the system is open to free $O_2$. For the charge disproportionation reaction model, the Gibbs energy is minimized as a function of $P$-$T$ such that the system is closed with respect to oxygen composition.

Eqs. M3 and M4 detail the phases and sublattice mixtures in a compact form for both models, while Eqs. M5 and M6 express the total thermodynamic potentials of the system. Eq. M5 represents the expression of the grand potential, which is minimized for the oxidation reaction model, where $\Delta n$ is the amount of the absorbed $O_2$. This expression makes use of the expression for the grand potential ($\Omega$) as $\Omega = G(Pv+Fp) - \Delta n \times \mu(O_2)$ where $G(Pv+Fp)$ is the total Gibbs energy of the solid phases, Pv and Fp. Equation S6 is the expression of the Gibbs energy, which is minimized for the chg. disp. reaction model.

We use an ideal solution model to formulate the Gibbs and grand potential energies in Eqs. M5 and M6, which treats the $Mg^{2+}$, $Si^{4+}$, $Al^{3+}$, $(Fe^{2+,HS})_{Fp}$, $(Fe^{2+,LS})_{Fp}$, $(Fe^{2+,HS})_{Pv}$, $(Fe^{3+,HS,A-site})_{Pv}$, $(Fe^{3+,HS,B-site})_{Pv}$, and $(Fe^{3+,LS,B-site})_{Pv}$ species as non-interacting species on each relevant sublattice. We have also checked the non-ideality effect and find that it doesn't have any qualitative impact on the result of our ideal solution model (the details are discussed in SI section 1.2 [*Blum and Zunger*, 2004]). If one considers these effects, quantitative changes are <0.09 for the Fe partitioning coefficient and <0.01 for the $Fe^{3+}/\Sigma Fe$ in Pv (the relative content of $Fe^{3+}$ in Pv) at all the lower mantle relevant *P-T* conditions. There is also no change for the Pv B-site occupancy. We thus did not consider the non-ideality effect further in our model. In the ideal solution approximation, the mixture enthalpies (*H*) can be written as a linear combination of their endmember values. The enthalpy of the endmembers in the system were obtained DFT calculations. The configurational entropy ($S_{config}$) and electronic-magnetic entropy of different Fe states ($S_{mag}$) are modeled in the ideal non-interacting system following the approaches in Refs. [*Tsuchiya et al.*, 2006; *Xu et al.*, 2015]. The only additional terms required in the present model vs. the previous reference [*Xu et al.*, 2015] are those in the configurational entropy for systems containing Al. These are treated in the standard way for a multicomponent ideal solution sublattice [*Chang and Oates*, 2009], e.g., as illustrated in Eq. M7 for the configurational entropy $S_{config}(Pv,B\text{-site})$ of the B-sublattice in Pv.

With the exception of a special situation for oxygen described next, no vibrational contributions to the free energies are included, as they are assumed to vary slightly between the different possible solid states of the system. The calculation of the vibrational effect is shown in SI section 1.3 [*Anderson*, 1989; *Hsu et al.*, 2010b; *Sha and Cohen*, 2010]. Similar to the non-ideality effect, the inclusion of the vibrational effect doesn't have a significantly qualitative impact on our results. The quantitative changes are <0.1 for the Fe partitioning coefficient and <0.05 for the $Fe^{3+}/\Sigma Fe$ in Pv at all the lower mantle relevant *P-T* conditions. There is also no observable change for the Pv B-site occupancy. Therefore, the vibrational effect is not further considered in our model. The one exception to this approach is in the oxidation reaction model, where $O_2$ can be absorbed into the system from the gas phase. In this case the vibrational contributions of oxygen atoms are present in the gas phase model and therefore should not be ignored in the ionic solid phase. Therefore, in our thermodynamic model we include an approximate solid phase vibrational free energy for oxygen, which in described in SI section 4.1 [*Lee et al.*, 2009; *Yagi*, 1978].

The $fO_2$ in different capsules as a function of *P-T* and the calculation of the effective chemical potential of $O_2$ can be found in the work of Xu et al. (Section 2.1.1 and SI section 1 for details) [*Xu et al.*, 2015]. Using the expressions in S5 and S6 for the thermodynamic potentials of the system, we have then minimized these potentials to reach thermodynamic equilibrium conditions at any given conditions of the possible variables. More specifically, for the chg. disp. model, in order to find stable mineral phases with equilibrium iron content and spin/valence states, the Gibbs energy, $G_{total}$[*Chg. Disp.*], is minimized with respect to ($n_1$:$n_6$, $x$, $x_1$:$x_4$, $y$, $y_1$) at each given *P-T* condition. A corresponding minimization is also done for $\Omega_{total}$[*Oxidation*].

$$System_{Oxidation} = (n_2 + n_3 + n_4 + n_5)(Mg_{\frac{n_2+n_3+n_4+n_5-x_2n_3-x_1n_2-x_3n_4-x_4n_5}{n_2+n_3+n_4+n_5}} Fe^{2+}_{\frac{x_1n_2}{n_2+n_3+n_4+n_5}} Fe^{3+,HS}_{\frac{x_3n_4+x_4n_5}{n_2+n_3+n_4+n_5}} Al^{3+}_{\frac{x_2n_3}{n_2+n_3+n_4+n_5}})$$
$$(Si_{\frac{n_2+n_3+n_4+n_5-x_2n_3-x_3n_4-x_4n_5}{n_2+n_3+n_4+n_5}} Fe^{3+,HS}_{\frac{x_4n_5y_1}{n_2+n_3+n_4+n_5}} Fe^{3+,LS}_{\frac{x_4n_5(1-y_1)}{n_2+n_3+n_4+n_5}} Al^{3+}_{\frac{x_2n_3+x_3n_4}{n_2+n_3+n_4+n_5}})O_3 + n_1(Mg_{1-x}Fe^{2+,HS}_{xy}Fe^{2+,LS}_{x(1-y)})O$$

(M3)

$$System_{ChgDisp} = System_{Oxidation} + n_6 Fe^0$$

(M4)

$$\Omega_{total}[Oxidation] = \{G_{total}[ChgDisp] - n_6 H[Fe^0] + n_6 TS(Fe^0)\} - \Delta n \mu(O_2)$$

(M5)

$$G_{total}[ChgDisp] = n_1 y H[(Mg_{1-x}Fe_x^{2+,HS})O] + n_1(1-y)H[(Mg_{1-x}Fe_x^{2+,LS})O] + n_2 H[(Mg_{1-x_1}Fe_{x_1}^{2+,HS})SiO_3]$$
$$+ n_3 H[(Mg_{1-x_2}Al_{x_2}^{3+})(Si_{1-x_2}Al_{x_2}^{3+})O_3] + n_4 H[(Mg_{1-x_3}Fe_{x_3}^{3+,HS})(Si_{1-x_3}Al_{x_3}^{3+})O_3] +$$
$$n_5 y_1 H[(Mg_{1-x_4}Fe_{x_4}^{3+,HS})(Si_{1-x_4}Fe_{x_4}^{3+,HS})O_3] + n_5(1-y_1)H[(Mg_{1-x_4}Fe_{x_4}^{3+,HS})(Si_{1-x_4}Fe_{x_4}^{3+,LS})O_3]$$
$$+ n_6 H[Fe^0] - n_1 xy TS_{mag}(Fe^{2+,HS}, Fp) - n_1 x(1-y)TS_{mag}(Fe^{2+,LS}, Fp) -$$
$$n_2 x_1 TS_{mag}(Fe^{2+,HS}, Pv, A-site) - (n_4 x_3 + n_5 x_4) TS_{mag}(Fe^{3+,HS}, Pv, A-site) -$$
$$n_5 x_4 y_1 TS_{mag}(Fe^{3+,HS}, Pv, B-site) - n_5 x_4 (1-y_1) TS_{mag}(Fe^{3+,LS}, Pv, B-site) - n_6 TS(Fe^0)$$
$$-(n_2 + n_3 + n_4 + n_5)TS_{config}(Pv, A-site) - (n_2 + n_3 + n_4 + n_5)TS_{config}(Pv, B-site)$$
$$-n_1 TS_{config}(Fp, A-site)$$

(M6)

$$S_{config}(Pv, B-site) = -k_B (X_{Si} \ln X_{Si} + X_{Fe,3+,HS} \ln X_{Fe,3+,HS} + X_{Fe,3+,LS} \ln X_{Fe,3+,LS} + X_{Al,3+} \ln X_{Al,3+})$$

(M7)

The mechanism for creating oxygen vacancies is not specifically considered in our modeling here as the oxygen vacancies have been predicted to be a very high energy defect (see our previous work [*Xu et al.*, 2015] for details).

### 4.3. Parameters for the DFT calculations

Our *ab-initio* calculations are performed using the Vienna *ab-initio* simulation package (VASP) based on DFT. The projector augmented wave (PAW) method [*Blochl*, 1994] is used for the effective potentials of all the atoms in the system. The PAW potentials we used included valence electrons $2p^6 3s^2$ for Mg, $3s^2 3p^2$ for Si, $2s^2 2p^4$ for O, $3p^6 3d^7 4s^1$ for Fe and $3s^2 3p^1$ for Al. A cutoff energy of 600 eV is used to ensure that the plane wave basis is large enough for the calculations to converge.

It is well known that the normal Local Density Approximation (LDA) and Generalized Gradient Approximation (GGA) functionals often provide inaccurate energetics for transition metal oxides, including oxides containing Fe [*L Wang et al.*, 2006]. All the calculations in this work are therefore performed with HSE06 hybrid functional [*Heyd et al.*, 2003; *Paier et al.*, 2006] as implemented in the VASP code. The fraction of the exact Hartree-Fock exchange functional is set to be 0.25 (i.e., we set AEXX=0.25 in the INCAR file). The HSE06 functional has been shown to yield significantly more accurate energetics for transition metal redox reactions than standard LDA or GGA techniques [*Chevrier et al.*, 2010]. All details concerning the *k* points and supercell setup information are shown in SI Table S3 [*Dubrovinsky et al.*, 2000]. The choices of the *k* point mesh yield a convergence of the total energy of a supercell within 1 meV/atom, while the structural relaxation is converged to less than $10^{-3}$ eV in the total energy, yielding the average forces between atoms to be about 0.01 eV/Å. The validation of our *ab-initio* calculations for these high-pressure phases in comparison with experimental equation of states (EOS) is shown in SI section 3.1 [*Birch*, 1986; *K. Catalli et al.*, 2011b; *K. Catalli et al.*, 2010; *Dubrovinsky et al.*, 2000; *Fiquet et al.*, 2000; *Jacobsen et al.*, 2002; *J. F. Lin et al.*, 2013; *J. F. Lin et al.*, 2005; *Lundin et al.*, 2008; *Mao et al.*, 2011; *Mathon et al.*, 2004; *Speziale et al.*, 2001]. It should be noted here that among all the DFT calculated energies of the endmembers in our thermodynamic model, two of them are obtained based on semi-empirical approaches. The first one is the enthalpy of metallic Fe as a function of pressures, where the experimental EOS parameters are used. The other one is the enthalpy of LS Fp $(Mg,Fe^{LS})O$ as a function of pressures, where a constant energy shift is applied to match with previous experimental and theoretical results of the spin transition region (please refer to SI section 3.1 for details [*J. F. Lin et al.*, 2013; *Tsuchiya et al.*, 2006]).

### 4.4. Density and Bulk Modulus of the Lower Mantle Phases

To calculate the density and bulk modulus of the lower mantle phases along an expected geotherm [*Brown and Shankland*, 1981], we have considered high temperature effects on the volume and bulk modulus of the solid endmembers in the system at high pressures. It should be noted here that the thermal expansion influence is not included in our Gibbs energy minimization calculation to find the equilibrium state of the system. After we get the different endmembers contents in equilibrium, we consider the high-temperature effects on volume and bulk modulus to correct our corresponding DFT values $V(P)^{DFT}$, isothermal $K_T(P)^{DFT}$. Then we calculate the adiabatic bulk modulus $K_S$ for the bulk sound velocity calculation. After we obtain the density and the adiabatic bulk modulus of each lower mantle phase, the Voigt-Reuss-Hill average is used to calculate the bulk modulus of the lower mantle [*Lu et al.*, 2013; *Murakami et al.*, 2012]. Please refer to the SI section 4.2 for the calculation details of the lower-mantle density, bulk modulus and bulk sound velocity.

**Acknowledgements:** D. Morgan and S. Xu acknowledge financial support from the United States National Science Foundation (EAR-0968685).  J. F. Lin acknowledges financial support from the United States National Science Foundation (EAR-1446946). Computations in this work benefitted from the use of the Extreme Science and Engineering Discovery Environment (XSEDE), which is supported by National Science Foundation grant number OCI-1053575. The authors appreciate all the previous works cited in the Supplemental Information. The data shown in the figures and the tables are available by contacting the corresponding author (D. Morgan: ddmorgan@wisc.edu) upon request.

**Figure Caption:**

**Fig. 1 Spin and valence states of iron and their relative fractions in the lower mantle ferropericlase (Fp) and bridgmanite (Pv) along an expected geotherm** [*Brown and Shankland*, 1981]. Solid symbols represent our theoretically-predicted results, while colored lines are fitted to the data to show trends. (a) Fractions of the high-spin $Fe^{2+}$ in Fp and high-spin B-site $Fe^{3+}$ in Pv are plotted as red and blue lines, respectively. (b) Relative fractions of iron in lower mantle Fp, Pv, and metallic iron with respect to the total amount of Fe in a pyrolitic compositional model. The B-site $Fe^{3+}$ content in Pv is significantly smaller ($\leq 10^{-6}$ mole fraction of Fe) than the contents of other Fe species and is not shown for clarity.

**Fig. 2 Partition coefficient ($K_D^{Pv-Fp}$) of iron and deviations in the density and bulk sound velocity due to the Fp $Fe^{2+}$ spin transition in a pyrolitic lower mantle composition along an expected geotherm**[*Brown and Shankland, 1981*]. (a) Red data points with fitted solid curve are the $K_D$ considering the spin transition $Fe^{2+}$ in Fp. Blue data points with fitted dashed line curve are the $K_D$ without the influence of this spin transition; $K_D$ values continue to increase with increasing pressure along the geotherm. (b) The deviation is defined as $(X'-X_{ref})/X_{ref}$, where $X'$ is the physical property (density ($\rho$) or bulk sound velocity ($V_\Phi$)) calculated using a pyrolitic model across the Fe spin transition along an expected geotherm[*Brown and Shankland, 1981*], $X_{ref}$ is the reference value without the spin transition. Black and red data points and fitted solid curves correspond to the density and bulk sound velocity deviations (%), respectively, from the reference profiles without the spin transition of $Fe^{2+}$ in Fp.

**Fig. 3 Comparison of the partition coefficient ($K_D$) of iron between Pv and Fp from experimental and theoretical results**. The orange color region represents our theoretical predicted range of $K_D$ in a pyrolite composition model. The lower limit (black line) of the orange region corresponds to the condition that $\log fO_2 = \log fO_2$[diamond-carbonate], the upper limit (black line) corresponds to the condition that $\log fO_2 = \log fO_2$[diamond-carbonate] + 2. The blue color region represents our predicted $K_D$ range having a bulk

composition of olivine (does not contain any $Al^{3+}$) under the same $fO_2$ range used in for the pyrolite composition. Experimental $K_D$ values in a pyrolite composition model (grey symbols) [*Irifune*, 1994; *Irifune et al.*, 2010; *Kesson et al.*, 1998; *Murakami et al.*, 2005; *Prescher et al.*, 2014; *Sinmyo and Hirose*, 2013; *Wood*, 2000] (dashed grey curve is obtained from [*Irifune et al.*, 2010]) and olivine composition (green symbols) [*Auzende et al.*, 2008; *Kobayashi et al.*, 2005; *Sakai et al.*, 2009; *Sinmyo et al.*, 2008a] are also plotted for comparison.

**Fig. 4 Variations of iron chemistry in a pyrolitic lower mantle model.** (a) The fractions of the total amount of Fe in atom % in Fp, Pv, and metallic Fe phase ($Fe^0$) as a function of depth are represented by different color regions. The color gradient represents the high-spin (HS) fraction of iron in both Fp and Pv phases as shown in the vertical bar on the right. In the Pv region, the dashed line separates the A-site $Fe^{3+}$ fraction and A-site $Fe^{2+}$ fraction in Pv. (b) The Al content in Pv (per 3 oxygen atoms) is shown by the blue solid line.

**Figure. 1**

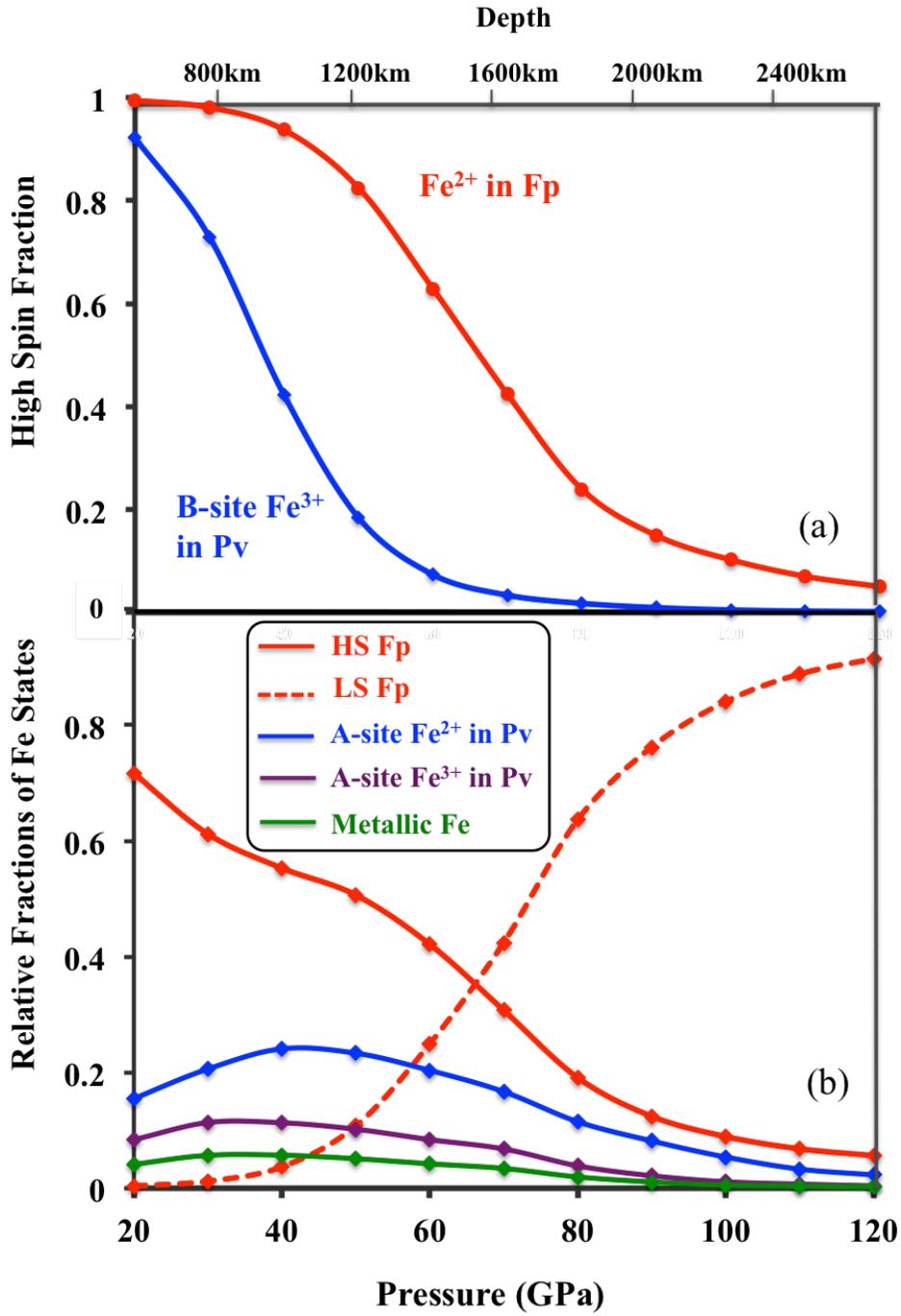

**Figure. 2**

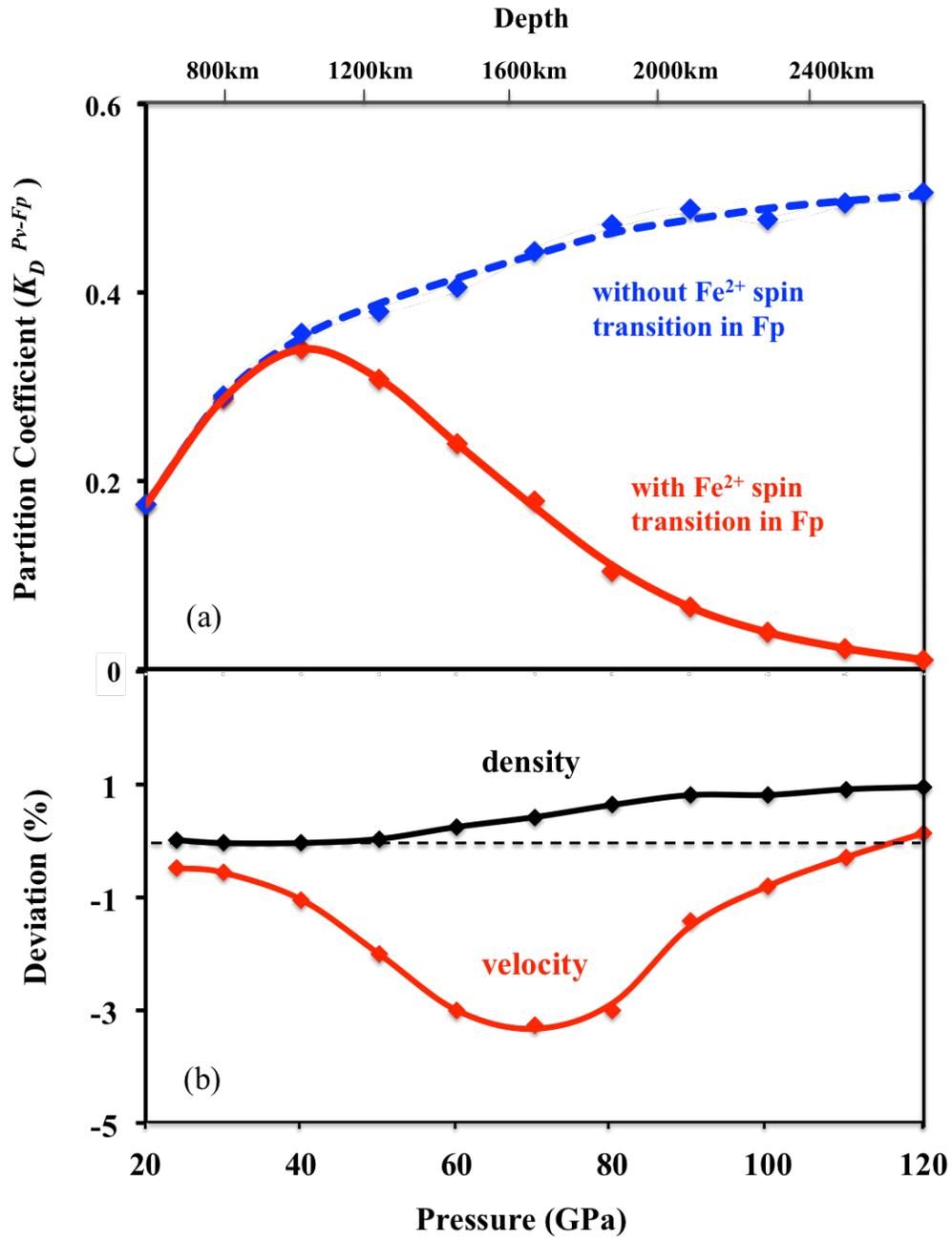

**Figure. 3**

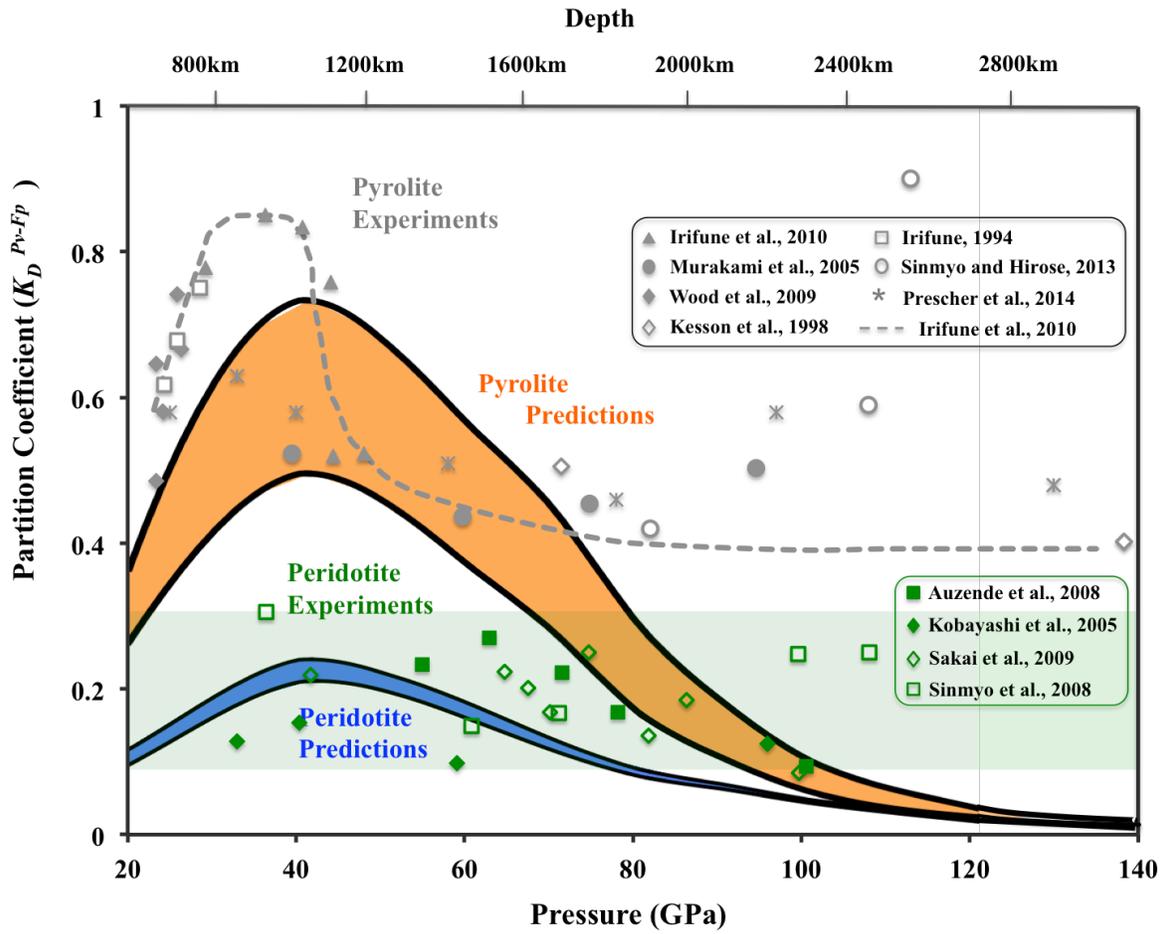

**Figure. 4**

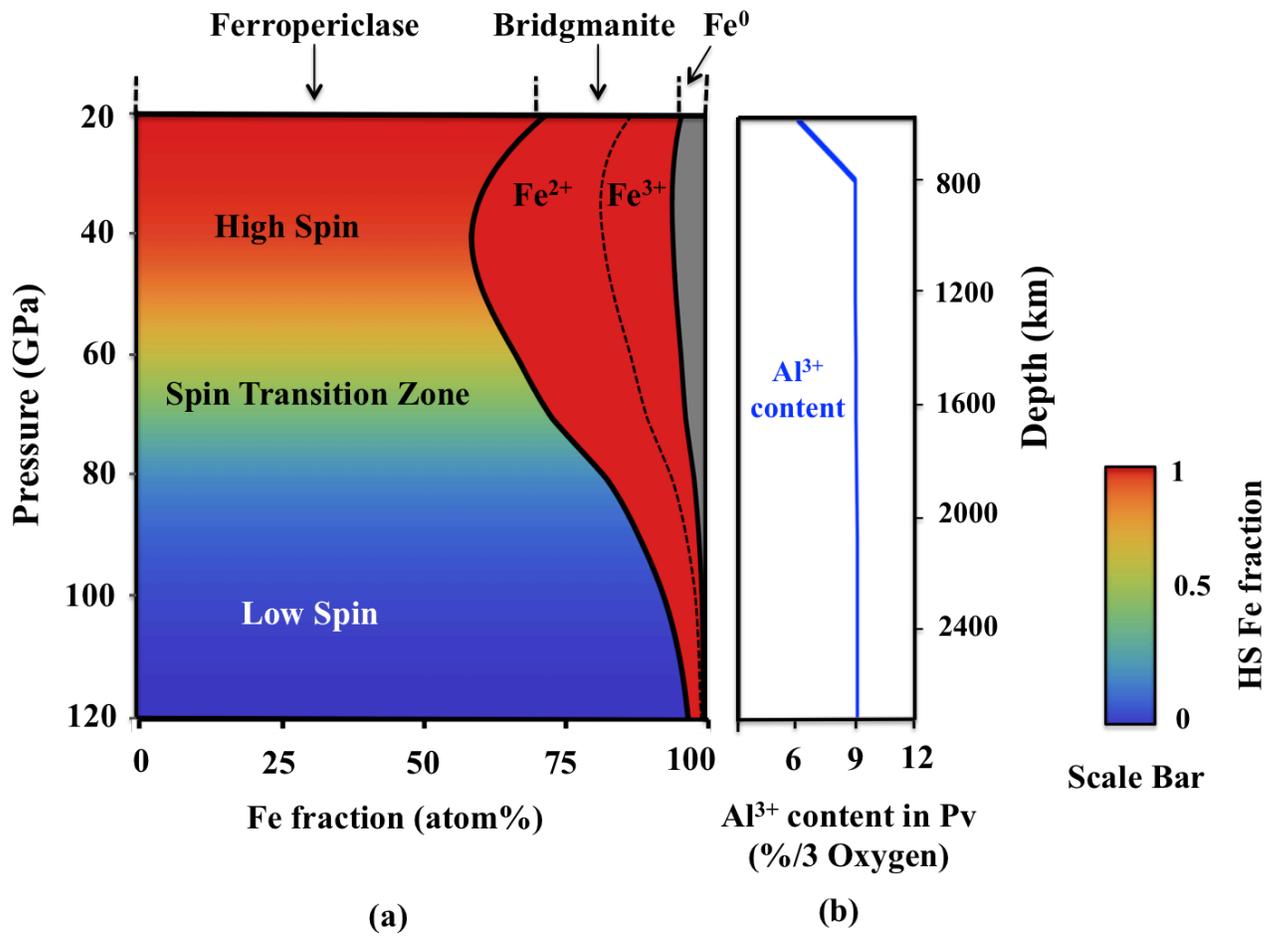